\documentclass{appolb}%
\usepackage{graphicx}
\usepackage{amsmath}

\def\vec#1{\ensuremath{\mathchoice
                     {\mbox{\boldmath$\displaystyle#1$}}
                     {\mbox{\boldmath$\textstyle#1$}}
                     {\mbox{\boldmath$\scriptstyle#1$}}
                     {\mbox{\boldmath$\scriptscriptstyle#1$}}}}

\begin{document}
%
%
\title{Phenomenological exploration of the strong coupling constant
in the perturbative and nonperturbative regions
}
\author{M. De Sanctis \footnote{mdesanctis@unal.edu.co}
\address{Universidad Nacional de Colombia, Bogot\'a, Colombia }
\\
}
\maketitle
\begin{abstract}
The  QCD running coupling costant is studied in the perturbative region, 
considering the existing experimental data, and also in the nonperurbative region,
 at low momentum transfer.
A continous phenomenological function is determined by means of three different models
also calculating the corresponding finite value of the vector quark self-energy.
These two quantities are used for
the vector interaction of a Dirac relativistic model for the charmonium 
spectrum.
The process required to fit the spectrum is  discussed and 
the relationship with  previous models 
is analyzed.

\end{abstract}
%
%
%
\section{Introduction}\label{intro}
In a series of previous works the author developed a Dirac relativistic
quark-antiquark  model
to study the spectrum of charmonium and, possibly, of other mesons.
The general structure of the quark vector interaction potential was introduced
in Ref. \cite{chromomds}.
In the same work the quark vector self-energy, denoted $\bar V_V$ in the present article, 
was studied,
determining its relationship with the  potential.

Successively, in Ref. \cite{localred},
considering the necessity of a relativistic study of the hadronic spectroscopy,   
the reduced Dirac-like equation (RDLE) of the model was introduced.
This equation is written  in the coordinate space in a local form.
%
An accurate calculation of the charmonium spectrum was then performed using a small
 number of free parameters in Ref. \cite{rednumb}.
Furthermore,
in a subsequent work \cite{relvar}, the Lorentz structure of the interaction terms was
studied in more detail, 
developing a covariant form of the  same RDLE.
The vector interaction used in those works
(aimed to the study of the charmonium spectrum)
 was essentially phenomenological,
consisting in a regularized Coulomb interaction where
the regularization was given by an effective structure of the interacting quarks.
No attempt was made  to establish a relationship with the perturbative
strong interaction given by QCD.
Those works also clearly showed  
that a vector interaction alone is not sufficient to give 
an accurate reproduction of the charmonium spectrum.
For this reason, the contribution of a
\textit{scalar interaction}  has been always included in the interaction of the RDLE.
In this respect,  the role of 
the scalar  interaction was studied in more detail 
in another work \cite{scalint},
also considering the possibility of using
a \textit{mass interaction}.
The results obtained with the two interactions were very similar.
In the same work the scalar and mass interactions have been tentatively
related to the excitation of the first scalar resonances of the hadronic spectrum.

%
\vskip 0.1 truecm
Finally, in the work \cite{effective}, the author started studying
a possible relationship between the vector interaction  for the charmonium spectrum
and the Quantum Chromo-Dynamics 
(QCD) \textit{effective running} strong coupling constant $\alpha_S(Q)$,
where the argument 
$Q$ represents  the standard quark vertex momentum transfer. 
In that work it was considered, for $\alpha_S(Q)$,
the \textit{effective charge}
extracted from the experimental data 
using the generalized Bjorken sum rule
\cite{Deur05, Deur08, Deur22}.
The properties of this quantity are  analyzed in detail
in the extensive reviews on  the QCD running coupling constant 
$\alpha_S(Q)$ \cite{Deurrev16,Deurrev23},  and
in the references quoted therein.
%
%
The results of our work \cite{effective} showed that  $\alpha_S(Q)$ 
can be parametrized by means 
of  a unique  function given by the  sum
of two terms: a gaussian function that dominates at low $Q$ and another function
that, at high $Q$, has the behavior predicted by the pertubative QCD,
in accordance with the experimental data.
This result can be considered \textit{encouraging} because it shows that,
in principle,  a unique function for $\alpha_S(Q)$ can be used in the perturbative and nonperturbative regions.

However, many issues remain unclear, justifying the development of the present exploration.
In more detail, the author makes the following criticisms:
\begin{itemize}
\item
To fit the charmonium spectrum, it has been necessary to introduce 
a general multiplicative constant, so that
$\alpha_S(0)\simeq 2$, not in agreement 
with the value of the effective charge at $Q=0$ given by Refs. \cite{Deur05, Deur08, Deur22}.

\item
In the same concern, one has to take into account that those experimentally extracted 
 data are referred to specific processes and  the \textit{extraction} procedure,
in the nonperturbative region,
may be process dependent.

\item
The  determination of the vector self-energy $\bar V_V$  was not analyzed 
in Ref. \cite{effective} and, in consequence, that quantity  was considered 
as a free parameter not related to the model  of the quark interaction.

\item
In Ref. \cite{effective} the quality of the fit of the charmonium spectrum 
was not optimal and should be improved.
\end{itemize} 

In consequence,
the main interest of the present investigation is to study
a general definition  of the strong coupling constant $\alpha_S(Q)$, with the aim of
 reproducing the hadronic spectroscopy
(in particular, charmonium spectrum)
  and, possibly,
improve the understanding of other
\textit{nonperturbative} hadronic phenomena,
such as quark confinement and  the emergence of hadronic mass.
In more detail, the low $Q$ behaviour of  $\alpha_S(Q)$,
mainly related to  hadronic spectra,
  must be \textit{matched}
with the high $Q$ behaviour, determined by QCD.
We highlight here that
an expression of
$\alpha_S(Q)$   ``in accordance'' with QCD and,
at the same time,
able to reproduce the quark interaction 
for the hadronic bound states
still represents a challenge for theoretical physics.

A crucially relevant property of $\alpha_S(Q)$
is that this quantity must not present a  divergence at low $Q$,
as it would be obtained applying incorrectly perturbative QCD.
On the contrary, $\alpha_S(Q)$  must go to a constant value as $Q \rightarrow 0$. 
%
%
In our previous works \cite{chromomds,rednumb,scalint}, 
this low $Q$ behaviour has been obtained 
by introducing an effective interacting quark structure, \textit{i.e.}
a form factor at the quark interaction vertex.
A relevant consequence of this procedure is to
\textit{regularize} the Coulombic potential
for $r\rightarrow 0$. 
A (different) technique of regularization has been shown to be beneficial  to avoid
unwanted singularities of the S-wave functions of charmonia at the origin \cite{coulreg},
when using another model of relativistic equation. 

Due to the interest
for a comparison with a very different theoretical approach, 
we recall that in  the Holografic Light-Front QCD \cite{Deurrev23} 
the $\alpha_S(Q)$  behaviour   at low $Q$ is  very similar to that 
given by  the form factors of our previous works \cite{chromomds,rednumb,scalint}.
In a recent work developed in the same theoretical framework
 \cite{hlfmatch}, the possibility of a smooth matching between the high $Q$ and low $Q$
behaviour of $\alpha_S(Q)$ is analyzed.
The authors find the matching point above $Q\simeq 2$ GeV, 
not very different  with respect to the results 
of the present work. 
At $Q=0$  they find, as in Refs. \cite{Deur05, Deur08, Deur22},
 $\alpha_S(0)\simeq \pi$ while our relativistic model
requires $\alpha_S(0)\simeq 2$.

Also, in the Richardson
model \cite{richar79},  a static potential for the 
constituent-quark interaction is introduced. 
This potential  grows linearly with
the quark distance.
From this potential one can formally obtain an effective coupling constant
that, however, is 
\textit{divergent} as $Q\rightarrow 0$.

We conclude this introduction pointing out that, in any case, further effects 
must be considered to understand completely the charmonium spectroscopy,
in particular for the high excitation states.
A semirelativistic screened potential model has been proposed for a comprehensive study 
of the mass spectrum and decay properties of charmonium states \cite{spectdec}.

Multiquark (virtual) states play a role of primary importance in the spectroscopy
of high energy states,
as discussed in Ref. \cite{santp} in the framework of the unquenched quark model.
The  exotic multiquark  states have been studied also considering three quark interactions
\cite{threebody}.
A very recent work proposes a comprehensive model of hadronic phenomenology
by using a Born-Oppenheimer effective theory to describe standard 
and exotic states \cite{bramb24}.
Investigation on hadronic phenomenology is still very active and no definitive conclusion 
has been obtained, suggesting the use different methods and techniques to understand
all the aspects of the problem.

\vskip 0.2 truecm

In this framework, the exploration performed in  this work is organized in the following way.
We first recall the  theoretical model on which the investigation is based.
In more detail, 
in the next Subsect. \ref{symbnot}, the notation and conventions are introduced and explained.
In Sect. \ref{dyn}, the dynamical model of our RDLE is summarized for the present study. 
In Sect. \ref{vectmcspace}, the vector interaction in coordinate space is derived from 
the corresponding momentum space expression, taking into account the form of $\alpha_S(Q)$.
The main properties of the vector self-energy $\bar V_V$ are analyzed, considering its relationship
with  $\alpha_S(Q)$.

We start our exploration  of $\alpha_S(Q)$ performing, in Sect. \ref{running},
 a general phenomenological survey of this quantity;
we consider the main issues of perturbative QCD at high $Q$ and the results  
of our previous works, concerning the charmonium spectrum, at low $Q$.

Then, taking into accont this survey, 
we study   $\alpha_S(Q)$ by introducing three models 
with an increasing level of complexity.
The results of each model are used as an ``input'' for the following one.
In particular,
in Sect. \ref{model1}, a simple model is introduced
by means of a piecewise function (with two intervals) for  $\alpha_S(Q)$.
We require continuity of $\alpha_S(Q)$ and of its first derivative at the matching point
$\bar Q$.

In Sect. \ref{model2}, $\alpha_S(Q)$ is represented by the sum of two functions:
the first one mainly takes into account the low $Q$ behaviour, while the second one
gives the high $Q$ behaviour by means of a regular expression inspired by the perturbative
QCD function.

We also study, in Sect \ref{model3}, a unique differential equation for $\alpha_S(Q)$
whose analytic (implicit) solution allows to represent the running coupling constant 
for all values of $Q$.
For the three models different strategies are studied to determine 
the \textit{finite value} of the vector quark self-energy.  

Finally, in Sect. \ref{charmspectr}, the obtained charmonium spectrum is displayed 
and discussed.
Its structure is similar to that of our previous works, showing that a general expression 
for $\alpha_S(Q)$ can be also  used to reproduce in detail the $c \bar c$ excitation states.
Some general comments are made and the provisional conclusions of this exploration are drawn. 


\subsection{Notation and conventions}\label{symbnot}

The following notation and conventions are used in the paper.

\begin{itemize}
\item  
The invariant product between four vectors is standardly written as: 
$V^\mu U_\mu= V^\mu U^\nu g_{\mu \nu}=V^0 U^0- \vec V \cdot \vec U$.
\item
The lower index $i =1,2$ represents the \textit{particle index},
referred to the quark ($q$) and to the antiquark ($\bar q$).
The generic word ``quark'' will be used for both particles.
\item
We shall use, for each quark,  the four Dirac matrices $\gamma_i^\mu$,
the three matrices $\vec \alpha_i= \gamma^0_i \vec \gamma_i $ 
and $\beta_i= \gamma^0_i$.
\item
The symbol $ q^\mu=(q^0, \vec q)$ will denote
the vertex 4-momentum transfer.
\item
We shall
neglect  the retardation contributions, setting $q^0=0$ for the time component
of the 4-momentum transfer.
This approximation is consistent with the use of the Center of Mass Reference Frame
for the study of the $q \bar q$ bound systems.
\item
In consequence, the \textit{positive} squared four momentum transfer $Q^2$
takes the form $Q^2=-q_\mu q^\mu= \vec q^2$, that is $Q=|\vec q|$.
As argument of the strong running coupling constant, we use the variable $Q$ 
and not $Q^2$, as it is frequently done.
For the calculations of Sect. \ref{model3} we shall introduce $t=Q^2$.
\item
The subindex $X$ will be used to denote, 
for the parameters $\bar V_X$ and $r_X$,
the scalar ($X=S$) or mass ($X=M$) character of the corresponding interaction.
\item
Finally, throughout the work, we use the standard natural units, 
that is $\hbar=c=1$.\\
\end{itemize}

\vskip 1.0 truecm
\section{ The dynamical model for the calculation of the charmonium spectrum}\label{dyn}
For completeness we recall here the main aspects of the dynamical model that is
used for the calculation of the charmonium spectrum \cite{localred,rednumb}.
The starting point of this model is
the Dirac-like equation for two interacting particles, 
that is written in the following form:
\begin{equation}\label{dle}
(D_1+D_2 +W)|\Psi>=0
\end{equation}
where we have used, for each quark, the Dirac operator
\begin{equation}\label{di}
 D_i=\vec \alpha_i \cdot \vec p_i +\beta_i m_i -E_i~,  
\end{equation}
The Dirac matrices have the standard definition given in Subsect. \ref{symbnot};
$\vec p_i$, $E_i$ and $m_i$ respectively represent the three-momentum,   energy 
and mass 
of the $i$-th quark; $W$ is the whole  interaction operator.
In  Eq. (\ref{dle})
the total operator  is applied to a two-particle Dirac state $|\Psi>$.\\
In Refs. \cite{localred,rednumb} we have introduced the reduction operators
\begin{equation}\label{defk}
K_i=
\begin{pmatrix} 1 \\ 
                {\frac {\vec \sigma_i \cdot \vec p_i} {m_i +E_i}  }
\end{pmatrix}
\end{equation}
that vinculate the lower and the upper components of the Dirac spinors.
By means of these operators, the RDLE is written as:
\begin{equation}\label{dir2red1}
 K_1^\dag \cdot K_2^\dag
(D_1 + D_2  +W )
       K_1
 \cdot K_2
| \Phi> =0
\end{equation}
where $| \Phi>$ is now a two-particle Pauli state.

This kind of reduction is particularly suitable for systems in which
a vector and a scalar interaction are present.
For the study of charmonium (and other $ q \bar q$ sistems),
we consider equal mass quarks setting $m_1=m_2=m_q$; 
we use the center of mass reference frame, where the total momentum is 
$\vec P= \vec p_1+ \vec p_2=0$.
In consequence, the quark momenta are 
$\vec p_2=-\vec p_1 =\vec p$, being $\vec p$ the quark relative momentum; 
its canonically conjugated operator represents the relative quark distance $\vec r$;
it means that in coordinate space one standardly has $\vec p=-i \vec \nabla$.
Furthermore, with the previous conditions, we can assume that the two quarks,
with equal masses, have the same energy:
$E_1=E_2=M/2$, where $M$ represents the total energy  of the system, 
that is the mass of the resonant state.\\
In this way, the RDLE of Eq. (\ref{dir2red1}) can be rewritten  in the following form
\begin{equation}\label{eq2redcm}
 \left[\left( 1 + {\frac {\vec p^2} {(M/2+m_q)^2} } \right)
\left( {\frac { 2 \vec p^2} {M/2+m_q} } +2m_q -M \right)
+ \hat W \right] | \Phi>=0
\end{equation}
where we have  introduced the reduced  quark interaction operator:

\begin{equation}\label{w2red}
\hat W= K_1^\dag \cdot K_2^\dag ~ W
~ K_1 \cdot K_2~.
\end{equation}
Eq. (\ref{eq2redcm}) represents a relativistic, energy-dependent and three-dimensional wave equation (in operator form) that is used to study, in this work, the charmonium spectrum.
It does not give rise to
spurious free solutions with vanishing total energy; in other words, 
it is not affected by the continuum dissolution desease.
This point and  other
formal properties of  Eq. (\ref{eq2redcm}),
in comparison to different relativistic equations,
 have been studied
in detail  in Ref. \cite{localred}.
Analogously to the Schr\"odinger equation, the \textit{ket} state $| \Phi>$
can be projected onto the \textit{bra} $ < \vec r | $, obtaining a coordinate 
space wave function. 
Recalling that the reduction operators $K_i$ of Eq. (\ref{defk}) are 
\textit{local}  operators,
we note that,
if the original, relativistic interaction $ W$ is local, 
also the reduced interaction $\hat W$, given by Eq. (\ref{w2red}),  is local and
Eq. (\ref{eq2redcm}), projected onto $ < \vec r | $,
gives a projected equation local overall.
If nonlocal terms were present in the original interaction $W$, 
also the reduced interaction $\hat W$ would be nonlocal. 
In this case $| \Phi>$
should be conveniently projected onto the momentum \textit{bra} $ < \vec p | $,
obtaining a momentum space wave function; in consequence,
Eq. (\ref{eq2redcm}) would give
an integral equation in the momentum space.
However, this second possibility is not considered in this work
where  we  take for the interaction $W$ a local expression.
As analyzed in Ref. \cite{relvar}, this choice is not in disagreement
with the relativistic character of the model.
The variational procedure used to solve Eq. (\ref{dir2red1}) is  discussed 
in Sect. 7 of Ref. \cite{localred}; in this way, the mass values of the resonant states of 
the charmonium spectrum are reliably determined.

As analyzed in the previous works, the interaction $W$ is given by a vector term $W_V$,
directly related to the QCD interaction, plus another term $W_X$ that must be introduced 
to reproduce accurately the charmonium spectrum. 
This term can be of scalar type ($X=S$) or of mass type ($X=M$). 
Both  interactions, used in the previous works, gave good results for the charmonium spectrum. 
Also in the present work we shall find that similar results are obtained in  the two cases.
For clarity, we give (in the center of mass reference frame)  
the Dirac structure of $W_V$, $W_S$ and $W_M$:

\begin{equation}\label{W_V}
W_V=\left[\bar V_V + V_V(r) \right]
\gamma_1^0 \gamma_2^0 \cdot
\gamma_1^\mu \gamma_2^\nu g_{\mu \nu}~,
\end{equation}

\begin{equation}\label{W_S}
W_S=V_S(r) \gamma^0_1\gamma^0_2 
\end{equation}
and
\begin{equation}\label{W_M}
W_M=V_M(r) \gamma^0_1\gamma^0_2 \cdot
 {\frac {1} {2} }
 (\gamma_1^0+\gamma_2^0) =
    V_M(r)
{\frac {1} {2} }
 (\gamma_1^0+\gamma_2^0)
\end{equation}
We note that the product $\gamma^0_1\gamma^0_2$ is present in all interactions listed
above because our RDLE is developed beginning from
the Dirac-like Eq. (\ref{dle}), that is written in the 
Hamiltonian form.
Considering the interaction operators of Eqs. (\ref{W_V}), (\ref{W_S}) and (\ref{W_M}),
we recall that
the corresponding reduced operators  $\hat W_V$ and $\hat W_S$ 
have been calculated in Eqs. (C.1) -(C.5) of Ref. \cite{localred};
the operator $\hat W_M$
has been given in Eq. (A.4) of Ref. \cite{scalint}.

The vector potential interaction function $V_V(r)$ of Eq. (\ref{W_V})
 will be analyzed for this work
in the next Sect. \ref{vectmcspace}.
The positive constant term $\bar V_V$ is, in any case, necessary to fit 
the charmonium spectrum and, for the case of phenomenological potentials
with quark form factors \cite{chromomds,rednumb},
represents the \textit{finite}  vector quark self-energy.
In the present work, this  quantity
will be calculated, in Sects. \ref{model1}, \ref{model2} and \ref{model3}, 
by discarding the infinite contribution
due to the high $Q$ behaviour of $\alpha_S(Q)$.
In Eqs. (\ref{W_S}) and (\ref{W_M}),
for the scalar or mass potential function $V_X(r)$, we take the same Gaussian expression 
of our previous works \cite{rednumb,scalint,effective}, that is
\begin{equation}\label{vxgauss}
V_X(r)=- \bar V_X \exp \left(- {\frac {r^2} {r_X^2}} \right)~.
\end{equation}
Furthermore, as in Refs.  \cite{rednumb,scalint}, $\bar V_X$ is determined by means of the following balance equation
\begin{equation}\label{balance}
\bar V_V= 2m_q- \bar V_X
\end{equation}
that expresses the equality between the positive vector self-energy $\bar V_V$
and the sum of the quark masses and the negative $X$-interaction self-energy
$-\bar V_X$.
Some comments about the values obtained for $\bar V_X$ will be given
in Sect. \ref{charmspectr}.

Finally, Eq. (\ref{eq2redcm}) will be solved with the same technique used in our previous works
\cite{localred,rednumb,relvar,scalint,effective}
by using the expressions of $V_V(r)$ and $\bar V_V$ that will be obtained in the models
of Sects. \ref{model1}, \ref{model2} and \ref{model3}.

\vskip 1.0 truecm
\section{ 
The vector interaction in momentum and coordinate space;
the quark vector self-energy
}\label{vectmcspace}
Even though this subject has been accurately discussed in Ref. \cite{effective},
for clarity and completeness, we analyze again here this point. 
Our RDLE of Eq. (\ref{eq2redcm}) 
has been formulated in the coordinate space. 
In order to introduce into this model the momentum dependent 
running coupling constant $\alpha_S(Q)$, 
it is strictly necessary to establish the connection 
between the expressions of the vector interaction potential written 
in coordinate space and in momentum space.

\vskip 0.2 truecm
\noindent
In this work we make the hypothesis that there exists a unique, nonsingular
function  $\alpha_S(Q)$ that represents both the QCD running 
of the strong coupling constant and the nonperturbative structure effects
at low momentum transfer.
By means of this quantity
the tree-level vector interaction in the momentum space,
for a $q \bar q$ system,
 can be written, in general, as
\begin{equation}\label{intmom}
{\cal W}_V(Q)= -{\frac 4 3} {\frac {4 \pi} {Q^2} } \alpha_S(Q)
\gamma_{1}^\mu \gamma_2^\nu g_{\mu \nu}
\end{equation}
where
$4/3$ represents   the color factor
in the $q \bar q$ case. 
Our model, in the present form, does not contain retardation effects,
that is we set $q^0=0$.

In this way,
performing the Fourier transform, one determines the corresponding expression 
in the coordinate space
\begin{equation}\label{fourtrans}
{\cal W}_V(r)=\int {\frac {d^3 q} {(2\pi)^3}}
 \exp{(i \vec q \cdot\vec r)} {\cal W}_V(Q)
\end{equation}
Multiplying the previous expression by $\gamma^0_1 \gamma ^0_2$
from the left, one has
\begin{equation}\label{v_vr}
\gamma^0_1 \gamma ^0_2{\cal W}_V(r)=
V_V(r) 
\gamma_1^0 \gamma_2^0 \cdot
\gamma_1^\mu \gamma_2^\nu g_{\mu \nu}
\end{equation}
that represents the  vector interaction 
$W^V$ introduced in Eq. (\ref{W_V}), without the
self-energy term $\bar V_V$. 
In particular, $V_V(r)$, that represents
 the vector  interaction potential in the coordinate space,
is given by the following Fourier transform
\begin{equation}\label{potfourtrans}
V_V(r)=  -{\frac 4 3} 
 \int {\frac {d^3 q} {(2\pi)^3}}
 \exp{(i \vec q \cdot\vec r)}
{\frac {4 \pi} {Q^2} }  \alpha_S (Q)~.\\
\end{equation}
%
Considering  the previous equation, 
in the first place, we recall that,
in the case of a \textit{constant} $\alpha_S (Q)$,
one would go back to a standard Coulombic interaction.
More precisely,  for $\alpha_S(Q)= \alpha_{Coul}$
 one would obtain in the coordinate space
a pure Coulombic $ q \bar q$ potential
\begin{equation}\label{purecoul}
V^{Coul}_{V }(r)=  -{\frac 4 3} {\frac  { \alpha_{Coul} }  {r}}~.
\end{equation}
In  the case of QED, for point-like particles,  the running of the coupling
constant is extremely  slow (and growing with $Q$).
For this reason  the use of a truly constant  quantity 
$\alpha_{em}$ is suitable for the study of many electromagnetic processes.
We also recall that the small value of $\alpha_{em}$ allows for a perturbative treatment
of the interaction. 
On the contrary, in QCD, as it will be discussed in the next Sect. \ref{running},
$\alpha_S(Q)$ decreases at large $Q$, giving rise to the \textit{asymptotic freedom} 
of the theory, but grows at small $Q$, invalidating in any case a perturbative approach. 
Furthermore, at phenomenological level,
the potential of Eq. (\ref{purecoul})  is not able to reproduce with good accuracy the charmonium spectrum,
also if other interactions with  different tensorial structures are introduced.

For this reason,
a model of the vector interaction was  
studied in Ref. \cite{chromomds} and then used in Ref. \cite{rednumb} to calculate 
the charmonium spectrum by means of the RDLE of Eq. (\ref{eq2redcm}).
In this model the quarks are considered as \textit{extended} particles
that give rise to  the chromo-electric field that, in turn, 
mediates the strong interacion.\\
Specifically, 
an \textit{accurate reproduction of the charmonium spectrum} 
has been obtained by using a Gaussian color charge distribution for each quark:
\begin{equation}\label{rhogauss}
\rho( x)={\frac {1} {(2 \pi d^2)^{3/2} }}
\exp\left(-{\frac {\vec x^2} {2d^2} } \right)~.
\end{equation}
From this distribution one obtains,
in the momentum space, the following vertex form factor
\begin{equation}\label{ffgauss}
F(Q)=\exp \left(-{\frac {Q^2 d^2} {2}} \right)
\end{equation}
Considering  one form factor for each quark vertex, one obtains
for the (effective) strong coupling constant
the following  nonperturbative expression, specific of our previous calculations:

\begin{equation}\label{alphagauss}
\alpha_S^{np}(Q)=  \alpha_V [F(Q)]^2= \alpha_V
\exp (- Q^2 d^2 )~.
\end{equation}
By setting
\begin{equation}\label{dtau}
d=1 / \tau
\end{equation}
one gets the standard expression that will be displayed in  Eq. (\ref{alpha_np})
of the following Sect. \ref{running}.
For completeness, we recall the numerical values of the parameters \cite{rednumb}:
\begin{equation}\label{numparnp}
 \alpha_V=1.864,~~  d= 0.1526 \text{ fm}, ~~ \tau=1.293 \text{ GeV}.
\end{equation}
The function $\alpha_S^{np}(Q)$ of Eq.(\ref{alphagauss}) has no singularities 
and  goes to the constant limit $\alpha_V$
as $Q\rightarrow 0$ but clearly its  high $Q$ behaviour 
is not in accordance with the experimental data that,
on the contrary,
are well reproduced by the $\alpha_S(Q)$ of perturbative QCD.

\vskip 0.5 truecm
\noindent
The Fourier transform defined in Eq. (\ref{potfourtrans}),
with $\alpha_S^{np}(Q) $ of Eq. (\ref{alphagauss}), can be performed analytically
giving the following
interaction potential 
\begin{equation}\label{vintgauss}
 V^{np}_V( r)=- {\frac 4 3}  {\frac {\alpha_V} {r}}
\text{erf} \left(  {\frac {r} {2d}  }\right)~.
\end{equation}
In Eq. (17) of Ref. \cite{rednumb}
the same result,
denoted there as $V^{int}(r)$,
was obtained by means of a different procedure
completely developed in the coordinate space.
For the following discussion,
note that the potential of Eq. (\ref{vintgauss}) is
\textit{regular} for $r \rightarrow 0$.
More precisely, we have:
\begin{equation}\label{vnp0}
V^{np}_V( 0)=
- {\frac 4 3}  {\frac {\alpha_V} {d}}
{\frac {1} {\sqrt{\pi}}}
\end{equation}

\vskip 0.5 truecm

We can now discuss another relevant  point of this work,
related to the quark \textit{self-energy}.
We recall that   a positive constant term,
 is often  introduced   phenomenologically,
as a \textit{free parameter},
 in the vector interaction of the
quark models to improve the reproduction of the experimental spectra. 
In Ref. \cite{chromomds} we showed that a nonpointlike color
charge distribution of the quarks gives rise to a positive
self-energy $\bar V_V$ that can be identified with the constant recalled above.
Furthermore we showed \cite{rednumb} that this constant has the value of
the vector interaction potential at $r=0$, with opposite sign:
\begin{equation}\label{vbarvdef}
\bar V_V=-V_V( 0)~.
\end{equation}
In consequence,
evaluating Eq. (\ref{potfourtrans}) at $r=0$, the following explicit expression
for the self-energy is obtained:
\begin{equation}\label{vbardefexpl}
  \bar V_V=  {\frac {8} {3 \pi}} \int_0^\infty dQ ~\alpha_S(Q)~.
\end{equation}
In this way $\bar V_V$ is not introduced as an additional free parameter
but is determined \textit{within the  model},
improving its consistency and  predictivity.\\ 
The total vector potential, obtained adding  $\bar V_V$
to the  interaction term, 
is vanishing at $r=0$ and
approaches the maximum value  $\bar V_V$ as $r\rightarrow \infty$.\\
Considering, as in Ref. \cite{rednumb}, 
the Gaussian $\alpha_S^{np}(Q)$ of Eq. (\ref{alphagauss}),
one can obtain the corresponding $\bar V_V$
 by means of  the integral of Eq. (\ref{vbardefexpl})
or, with Eq. (\ref{vbarvdef}), simply changing the sign in Eq. (\ref{vnp0}). 
We rewrite   the result, using Eq. (\ref{dtau}):
\begin{equation}\label{vbargauss}
\bar V_V= 
{\frac 4 3}  \alpha_V 
{\frac {\tau} {\sqrt{\pi}}}~.
\end{equation}
The numerical value for the self-energy, obtained in  Ref. \cite{rednumb},
was $\bar V_V= 1.813 ~  $ GeV.
%
The total vector potential function $ \bar V_V + V^{np}_V(r) $,
with $\bar V_V$ given by Eq. (\ref{vbargauss}),
was used successfully in Ref. \cite{rednumb}
to obtain an accurate reproduction of the charmonium spectrum.

For the case of a general (phenomenological) $\alpha_S(Q)$, 
the procedure recalled above is possible only if
the integral of Eq. (\ref{vbardefexpl}) is convergent
or, equivalently, if
 $V_V(0)$ is a \textit{finite}, 
negative quantity. 
Obviously, that procedure does not work for the pure Coulombic case
in which  $\alpha_{Coul}$ of Eq. (\ref{purecoul})   is a  constant quantity.
Considering 
the high $Q$ behaviour of the QCD coupling constant
(see $\alpha_S(Q)=\alpha_S^p(Q)$
of the following Eq.(\ref{alpha_pert}))
we note that this asymptotic  behaviour
does not allow for convergence of the integral of Eq. (\ref{vbardefexpl}).
For this reason, different strategies will be studied in the next sections
to avoid this inconsistency.
These strategies essentially consist in isolating in the 
$\alpha_S(Q)$ of the model a ``nonperturbative" part,
related to low values of $Q$,
that gives the correct self-energy $\bar V_V$; 
at the same time, we make the hypothesis that 
the infinite contribution, given by
the ``remainder" of  $\alpha_S(Q)$ and
related to the  perturbative running coupling constant at high $Q$,
can be discarded, being not observed in the physical interaction.
\vskip 1.0 truecm
\section{Introductory survey of the running coupling constant $\alpha_S(Q)$  }\label{running}

As starting point, we recall that in the  high $Q$ region,
the momentum dependence of $\alpha_S(Q)$ is given by
the perturbative running of the QCD coupling constant.
In other words,
in this region  $\alpha_S(Q)$  can be completely identified with the running coupling constant of QCD.
As we shall see in the following, the situation is much less clear at low momentum
where, in any case, it is necessary to construct a theoretical \textit{model}
for $\alpha_S(Q)$.
 \vskip 0.5 truecm
\noindent
In the first place,
we focus our attention on
 $\alpha_S(Q)$ as obtained from the experimental data  shown in Ref. \cite{pdg24}.
These data are given for $Q> 2.0$ GeV where the perturbative expansion 
of QCD is assumed to hold.
In particular, in this work, for the strong coupling constant
  we consider the  ``reference'' value given by
the current Particle Data Group average \cite{pdg24}:
\begin{equation}\label{alphamz} 
\alpha_S(M_Z)=0.118 ~~~~\text{with}~~~~ M_Z=91.1876~ \text{GeV}.
\end{equation}
For our analysis,
we try to reproduce the
experimental data by using a function inspired by
the standard QCD perturbative 
calculations at the leading order, as given, for example in Refs. \cite{pdg24,prs,chk}:
\begin{equation}\label{alpha_pert}
\alpha_S^{p}(Q)= {\frac {1}  { b \ln \left( {\frac {Q^2} {\Lambda^2}} \right) } }
\end{equation}
where $\Lambda $ represents the overall scale for the interaction \cite{prs}
and $b$ (also denoted as $b_0$ or $\beta_0$) can be related, in perturbative QCD, 
to the number of  active flavours $n_f$ by the following expression \cite{pdg24}: 
\begin{equation}\label{b}
b={\frac {33- 2n_f}  {12 \pi}  }~.
\end{equation}
Eq. (\ref{alpha_pert}) is obtained as solution of the renormalization group 
differential equation at the leading order. 
Some more details will be given in Sect. \ref{model3}.
In pertubative QCD,
Eq. (\ref{alpha_pert}), that takes into account the one loop contributions,
characterizes the asymptotic freedom of the theory.
Higher order  contributions, corresponding to two loops, three loops, etc.,
have been carefully calculated as discussed in the works \cite{pdg24,prs,chk}
and in the references quoted therein.
The effects related to these terms will not be taken into account 
in this phenomenological study because they
should  not be very relevant for the calculation of the  charmonium spectrum, 
at least at the present level of precision.

The function $\alpha_S^{p}(Q)$ of Eq. (\ref{alpha_pert})
cannot be used at low $Q$, say for  $Q< 2.0$ GeV.
In fact, this function 
\textit{grows} when decreasing $Q$
and becomes \textit{singular} for $Q=\Lambda$ \cite{pdg24,prs,chk}.
This behaviour is incompatible with the perturbative approach 
that was used to derive Eq. (\ref{alpha_pert}).
For our study, it does not allow to construct an interaction for the
calculation of the charmonium spectrum.
Additionally, we anticipate that
we shall  introduce in Sect. \ref{model2} a regularized form of  the function
of Eq. (\ref{alpha_pert}) that will be used for
the Model II of this article.

For the following developments, we need to determine,
in a phenomenological way,  
 the values of 
the two parameters $b$ and $\Lambda$ of Eq. (\ref{alpha_pert}).
To this aim,
we take:
\begin{itemize}
\item 
 the numerical values of Eq. (\ref{alphamz});
\item
the values
\begin{equation}\label{alphaq1} 
\alpha_S(Q_1)=0.215 ~~~~\text{with}~~~~ Q_1=5.0 ~ \text{GeV},
\end{equation}
given by the data of Ref. \cite{pdg24}.
\end{itemize}
With standard algebra we calculate  the two parameters, 
obtaining:
\begin{equation}\label{beta_lambda}
b=0.65842,~~~~ \Lambda=0.14622 ~ \text{GeV}~.
\end{equation}
By using Eq. (\ref{b}), one can determine the
\textit{effective} number of flavours for our phenomenological expression, obtaining
$n_f=4.089 ~.$

In the low $Q$ region, the nonperturbative effects are dominant and
no  \textit{universal} measurement of $\alpha_S(Q)$ exists, 
even though extractions of the data for \textit{specific} processes 
have been proposed, as discussed in Sect.\ref{intro}.\\
For  the present work we recall that $\alpha_S(Q)$ in this region  
is particularly relevant for the determination of the charmonium spectrum.
For this reason we take, \textit{indicatively}, as an example to start our survey, 
the phenomenological  results of our previous  works, in particular
Ref. \cite{rednumb}.\\
As it has been  explained
in Eqs.  (\ref{rhogauss}) - (\ref{dtau}) 
of the previous Sect. \ref{vectmcspace},
these results (transformed to momentum space) give rise to 
the following Gaussian expression
 \begin{equation}\label{alpha_np}
\alpha_S^{np}(Q)=\alpha_V \exp\left(- {\frac {Q^2} {\tau^2}} \right)
\end{equation}
where $\tau$  represents the scale of the strong
interaction in the nonpertubative region.
The numerical values of the parameters $\alpha_V$ and $\tau$ have been
recalled in Eq. (\ref{numparnp}).
In our previous works \cite{rednumb,scalint}, 
to reproduce the charmonium spectrum, 
we need  $\alpha_V\simeq 2$ 
 that is not compatible with the determinations 
of Refs. \cite{Deur05, Deur08, Deur22,hlfmatch}.
The physical meaning of Eq. (\ref{alpha_np}) 
can be related to an effective internal structure of the quarks \cite{rednumb,scalint},
as it has been explained in Sect. \ref{vectmcspace}.

We point out that, obviously, the complete $\alpha_S(Q)$ 
cannot be obtained by multiplying $\alpha_S^{np}(Q)$
(proportional to the product of two standard vertex form factors) 
by a regularized
perturbative running coupling constant because, at high $Q$, 
the faster decay of the form factors would \textit{obliterate} the slower
logarithmic decay of perturbative QCD.
On the contrary, the high $Q$ perturbative behaviour
and the low $Q$ nonperturbative behaviour of  $\alpha_S(Q)$ 
must coexist in the same function.
Tentatively, we can say that the quarks ``partly" interact 
perturbatively as point-like particles and ``partly"
as particles with an internal structure.

\begin{figure}
{
  \includegraphics{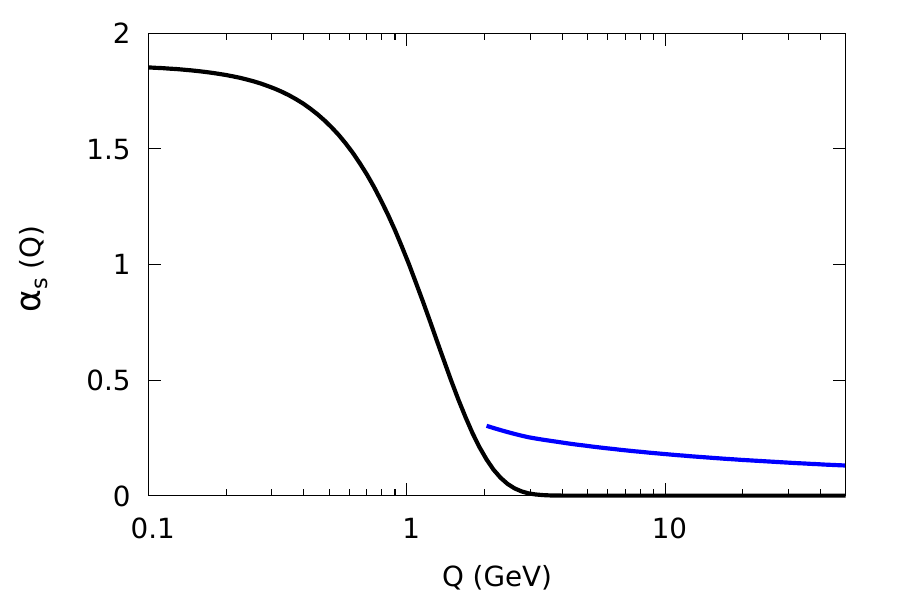}
}
\vspace{0.01cm}
\caption{
The black line represents an example of phenomenological nonperturbative 
strong coupling constant, that is
$ \alpha_S^{np}(Q)$ of  Eq. (\ref{alpha_np})
with the numerical vaues of the parameters given in Eq. (\ref{numparnp});
the blue line represents
 $\alpha_S^{p}(Q)$ of Eq. (\ref{alpha_pert}) that describes
 the experimental data in the perturbative region with
the values of the parameters given in Eq. (\ref{beta_lambda}).
The transition between the nonperturbative and perturbative regime is around
$Q \simeq 2.0$ GeV.
}
\label{fig1}       
\end{figure}

For the reader's orientation, we plot in Fig. \ref{fig1} 
the example of $\alpha_S^{np}(Q)$ given in Eq. (\ref{alpha_np})  and 
$\alpha_S^{p}(Q)$, as functions of $Q$.
The reader can appreciate that, as expected, the transition between the
two regions is found around $Q\simeq 2$ GeV.

In  this work we shall try to construct a unique, regular function  $\alpha_S(Q)$. 
Roughly speaking, we have to match the two curves Fig. \ref{fig1}.
In this respect we anticipate that much care must be exercised 
when performing this procedure
because the charmonium spectrum is extremely sensitive 
to the form of the interaction that is obtained
by means of this matching.

\vskip 4.0 truecm

\section{Model I. Piecewise function}\label{model1}
In this section, as a starting point of our exploration, 
we introduce a very simple definition of a
(unique) strong coupling constant as a function of the momentum transfer $Q$.
In this Model I, we consider a \textit{piecewise function} of $Q$ with two
intervals.
We make the hypothesis that in 
the first interval, defined by $ 0 \leq Q < \bar Q$,
$\alpha_S(Q)$ is mostly related 
to the nonpertubative effects of the quark interaction.
On the other hand, we assume that
in the second interval, 
defined  ~~ by $ Q \geq \bar Q$, $\alpha_S(Q)$
is given by  the perturbative form of the
QCD interaction. 
We require the \textit{continuity} of  $\alpha_S(Q)$ in $\bar Q$
and also the \textit{continuity of its first derivative}, in 
the same point.
The definition of the piecewise $\alpha_S(Q)$ function
of the Model I is given
by the following equation: 
\begin{equation}\label{alpha_1}
 \alpha_S(Q)= \begin{cases}
 \alpha_V-\alpha_G + \alpha_G \exp\left(- {\frac {Q^2} {\tau^2}} \right) \text{for} ~~0 \leq Q < \bar Q,\\
\alpha_S^p( Q)
~\text{for}~~ Q \geq \bar Q ~.
\end{cases}
\end{equation}
As discussed above, the definition of the coupling constant function 
in the second interval is given \textit{exactly}
by the standard $\alpha_S^p(Q)$ of Eq. (\ref{alpha_pert}). 
Moreover, the parameters $b$ and $\Lambda$ are fixed and
have the same numerical  values of Eq. (\ref{beta_lambda}).
In the first interval, as suggested by our previous studies \cite{rednumb,scalint}, 
we have a Gaussian term, proportional to $\alpha_G$;
furthermore we have to add the constant contribution $\alpha_V-\alpha_G$.
Note that, in this way, we have $\alpha_S(0)= \alpha_V$. 
The form of Eq. (\ref{alpha_1}) has been chosen in order to reproduce 
the experimental data of the running coupling constant 
and to fit the charmonium spectrum.

The conditions of continuity of the function  
and of the first derivative,
both calculated in $\bar Q$, give, respectively, 
the following equations:

\begin{equation}\label{cond1}
\alpha_V-\alpha_G + \alpha_G \exp\left(- {\frac {{\bar Q}^2} {\tau^2}} \right)=
\alpha_S^p({\bar Q})
\end{equation}
\begin{equation}\label{cond2}
-2 \alpha_G {\frac {\bar Q} {\tau^2}} \exp\left(- {\frac {{\bar Q}^2} {\tau^2}} \right)=
\alpha_S'^p({\bar Q})
\end{equation}
where
\begin{equation}\label{alphanpder}
\alpha_S'^p({\bar Q})=
- ~{\frac {2 } { b {\bar Q} \ln^2 \left( {\frac {{\bar Q} ^2} {\Lambda^2}}  \right) } }
\end{equation}
represents the first derivative of $\alpha_S^p( Q) $ calculated at $Q=\bar Q$.

\begin{figure}
{
  \includegraphics{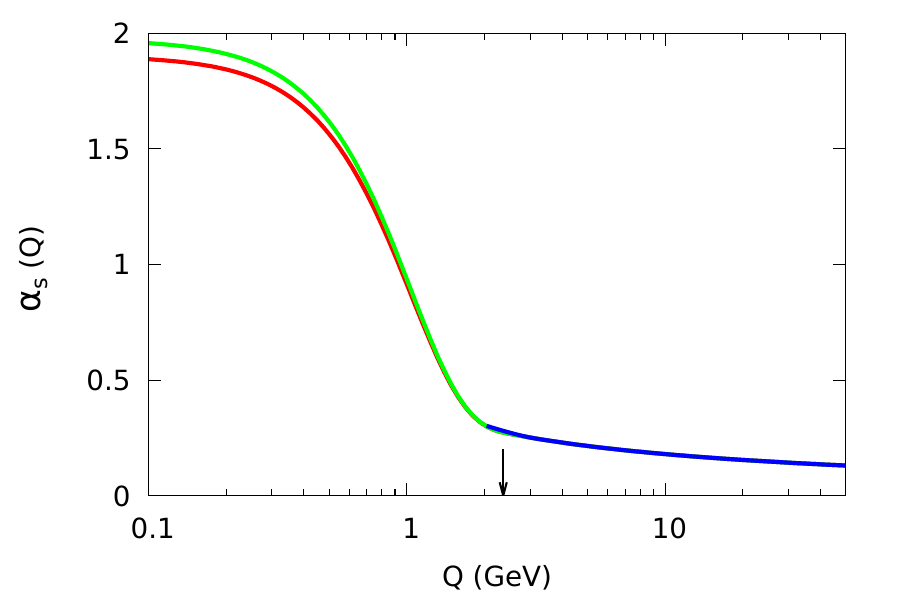}
}
\vspace{0.01cm}
\caption{
We plot $\alpha_S(Q)$ of Model I,
defined in Eq. (\ref{alpha_1}) with the conditions 
of Eqs. (\ref{cond1}) and (\ref{cond2}).
To fit the charmonium spectrum we have used the scalar interaction
with the parameters of the column (I S) of Tab. \ref{tabpar12};
in this case $\alpha_S(Q)$ is plotted with a red line.
For the case of the mass interaction,
with the parameters of the column (II S) of Tab. \ref{tabpar12},
$\alpha_S(Q)$ is plotted with a green line.
The blue line represents the perturbative function $\alpha_S(Q) $
of Eq. (\ref{alpha_pert}), with the values of Eq. (\ref{beta_lambda}),
 that fits the experimental data.
The  arrow indicates the values of $\bar Q$ for the scalar and mass interaction. 
As given in Tab. \ref{tabpar12} and discussed in Sect. \ref{model1},
these two values are very close and are not distinguishable 
in this graphic.
}
\label{fig2}       
\end{figure}

In order to obtain the \textit{finite} quark self-energy $\bar V_V$,
we take the definition of Eq. (\ref{vbardefexpl}) but perform the integral
only up to $\bar Q$,
disregarding the infinite contribution that would be obtained
integrating  from $\bar Q$ to $\infty$.
It means that, for this Model I we have
\begin{equation}\label{vbardefmod1}
  \bar V_V=  {\frac {8} {3 \pi}} \int_0^{\bar Q} dQ ~\alpha_S(Q)~.
\end{equation}
By means of the definition given in Eq. (\ref{alpha_1}),
this integral can be calculated analytically,
obtaining the following expression:
\begin{equation}\label{vbarmod1}
\bar V_V=  {\frac {8} {3 \pi}}\left[(\alpha_V-\alpha_G)\bar Q +
  \alpha_G {\frac  {\sqrt{\pi}} {2} } \tau \cdot \text{erf} \left({\frac {\bar Q} {\tau}} 
	\right) \right]~.
	\end{equation}
%
%
We note that in Model I we have four parameters:
$\alpha_V,~\alpha_G,~\tau$ and $\bar Q$ that are
vinculated by
the two conditions of Eqs. (\ref{cond1}) and (\ref{cond2}).\\
In Fig. \ref{fig2} we plot $ \alpha_S(Q)$ of the Model I.
The charmonium spectrum is reproduced by using the scalar interaction and 
the mass interaction.
The numerical values of the parameters, determined by the fit procedure, 
are shown respectively for the two interactions in the Column I S and I M
of Tab. \ref{tabpar12}. 
We note that $\alpha_V$ is slightly bigger for case of the scalar interaction.
The values of the constant $\tau$ of the Gaussian function are very similar
in the two cases. 
Also the values of $\bar Q$ (that separates the perturbative and nonperturbative interval)
are very similar for the scalar and the mass interaction.
The quality of the fit of the spectrum, given by the parameter $\Theta$,
that will be defined in Eq. (\ref{qual}), are displayed
at the bottom of Tab. \ref{tabpar12}, showing that a better reproduction
is obtained with the scalar interaction.

In summary,
by means of Model I we have learned that a two interval piecewise function 
(constructed requiring continuity and first derivative continuity)  
can represent $\alpha_S(Q)$ for all the values of the momentum transfer $Q$.
The value $\bar Q \simeq 2.362~ GeV$ that separates the two intervals, 
is also used as integration limit to calculate the finite value of $\bar V_V$. 
In the first interval it was necessary to use a Gaussian function plus a constant term.

\section{Model II. One analytic  function with the same definition 
for all the values of $Q$}\label{model2}
In this model we try to  represent the strong coupling constant as an analytic function 
with the same definition along the whole $Q$ axis.
The simplest way to construct this function
 is to use 
the sum of a Gaussian function for the nonperturbative region
``plus" a function that describes the perturbative behaviour of $\alpha_S(Q)$
at high $Q$.
However,
for the latter function it is not possible to take \textit{directly} $\alpha_S^p(Q)$ 
of Eq.  (\ref{alpha_pert}) because this function presents a singularity at $Q=\Lambda$.
To ``cure" this singularity we add the constant term $c_L>1$  to $ Q^2/\Lambda^2$ in 
 the argument of the logarithm.
We denote this new function as $\xi_L(Q)$.
In this way, we define $\alpha_S(Q)$ for the Model II, in the following form:
\begin{equation}\label{alpha_2}
\alpha_S(Q)=(\alpha_V-\alpha_L)\exp\left(- {\frac {Q^2} {\tau^2}} \right) 
+ \xi_L(Q) 
\end{equation}
with
\begin{equation}\label{xi}
 \xi_L(Q) =
 {\frac {1} { b \ln \left( c_{_L} +
{\frac {Q^2} {\Lambda^2}} \right) } }~.
\end{equation}
Furthermore,  we define $c_L$ as:
\begin{equation}\label{c_L_def}
c_{_L}=\exp \left({\frac {1} {b ~\alpha_L} } \right)~.
\end{equation}
Note that, by means of the previous definition,
 $c_{_L}$  \textit{does  not represent a new free parameter};
 furthermore its form has been chosen in
such way that,  in Eq. (\ref{xi}),  $\xi_L(0) = \alpha_L$;
in consequence, for the strong coupling constant of Model II, given by 
Eq. (\ref{alpha_2}),
we have $\alpha_S(0) = \alpha_V$.
Finally, the numerical value of $b$ given in Eq. (\ref{beta_lambda}) 
and the value of $\alpha_L$
that will be obtained in the following, numerically
give $c_{_L}>1$ so that, as discussed above,
 the function  $\xi_L(Q)$ does not present any singularity. 
We conclude this discussion recalling that the parametrization
of $\alpha_S(Q)$ given in Eq. (\ref{alpha_2}) is similar to that adopted 
in Ref. \cite{effective}.

\vskip 0.5 truecm
\noindent
To calculate the self-energy of the vector interaction $ \bar V_V $ 
in Model II,
we propose two different techniques, denoted as Technique A and Technique B.
For the Technique A, we note that at high $Q$
(where the contribution of the Gaussian term is negligible),
$\alpha_S(Q)$ of Eq. (\ref{alpha_2}), due to the presence of $c_L$, 
is slightly \textit{less than} $\alpha_S^p(Q)$ of Eq. (\ref{alpha_pert});
on the contrary,
decreasing $Q$, $\alpha_S(Q)$ becomes \textit{greather than} $\alpha_S^p(Q)$.\\
We can define, for the Technique A,
 $\bar Q$ as the value of momentum transfer for which the following
equality is satisfied:
\begin{equation}\label{qbardef}
\alpha_S(\bar Q)=\alpha_S^p(\bar Q)~.
\end{equation}
By using the expressions given in
Eq. (\ref{alpha_2}) and in Eq.  (\ref{alpha_pert}),
the value of $\bar Q$ can be determined numerically.

\begin{figure}
{
  \includegraphics{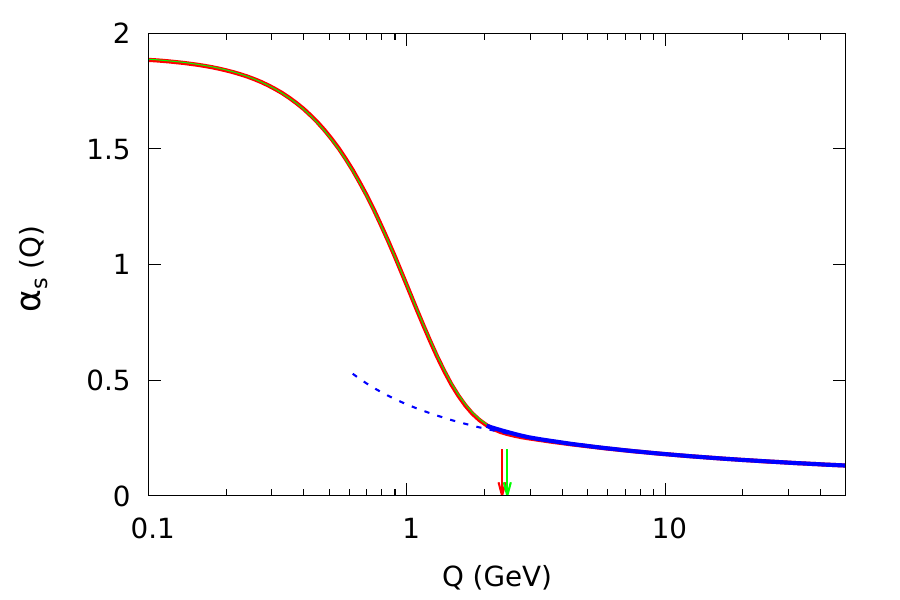}
}
\vspace{0.01cm}
\caption{ We plot $\alpha_S(Q)$ of
Model II A (see Eqs. (\ref{alpha_2}) -(\ref{c_L_def}))  
with the parameters of the scalar and mass interaction;
a red and a green line are used for the two interactions, respectively.
The two lines are partly superimposed.
 See Tab. \ref{tabpar12}, columns (II A S)
and (II A M) for the values of the parameters in the two cases.
The  blue line represents $\alpha_S^p(Q)$.
The dotted blue line represents this last function at low $Q$ and is plotted to illustrate
the method used to find $\bar Q$.
The values of this last quantity are represented by a red arrow and by a green arrow
for the scalar and mass interaction, respectively. 
}
\label{fig3}       
\end{figure}

Then, for the calculation of the self-energy we can use the same
expression used for Model I, 
that is Eq. (\ref{vbardefmod1}), but
integrating $\alpha_S(Q)$ of Eq. (\ref{alpha_2})
up to  the value $\bar Q$  determined by Eq. (\ref{qbardef}).
In this way the finite self-energy is given  by the nonperturbative effects
that dominate at $Q\leq \bar Q$
while the infinite contribution given by the
integration from $\bar Q$  to infinity  is discarded.\\
The numerical results of the parameters that reproduce
the experimental data of the strong coupling constant and fit 
the charmonium spectrum
 are shown in column (II A S) and
(II A M) of Tab. \ref{tabpar12} for the scalar and mass interaction, respectively.
The behaviour of $\alpha_S(Q)$ is displayed in Fig. \ref{fig3} where
the procedure discussed above for finding $\bar Q$ is also illustrated.

\begin{figure}
{
  \includegraphics{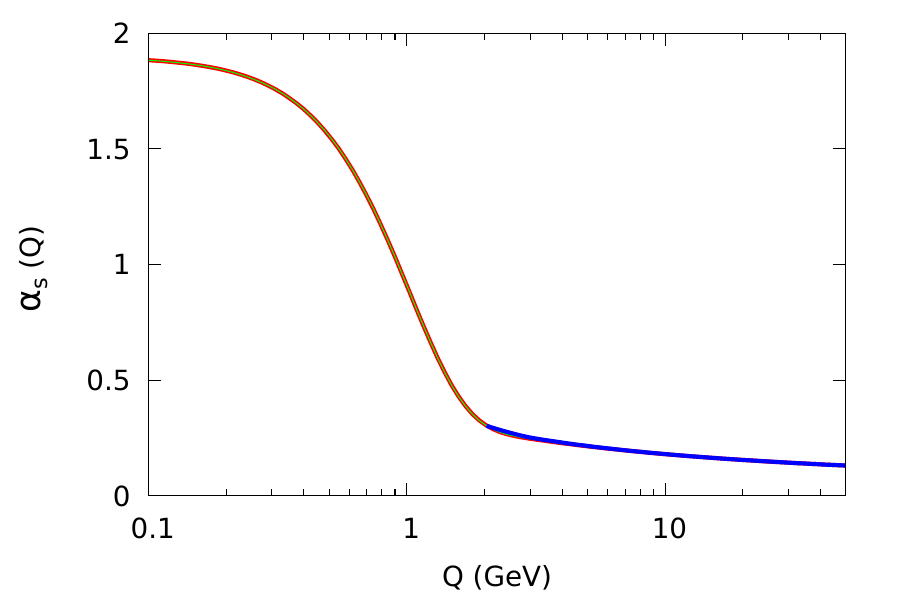}
}
\vspace{0.01cm}
\caption{ We plot $\alpha_S(Q)$ of
Model II B for the scalar and mass interaction.
The definitions and  conventions are the same as in Fig. \ref{fig3}.
}
\label{fig4}       
\end{figure}

\vskip 0.5 truecm
In order to introduce the Technique B, we consider,
in the first place, that the Gaussian function
(that is the first term of $\alpha_S(Q)$ in Eq. (\ref{alpha_2}) ) 
can be integrated up to infinity and
gives a standard, finite contribution to the self-energy.
This contribution can be calculated  
as in Eq. (\ref{vbargauss}) but replacing $\alpha_V $ with $\alpha_V - \alpha_G$.
However, also  the second term of Eq. (\ref{alpha_2}),
that is $\xi_L(Q)$,
``contains" some 
nonperturbative effects that give a contribution to the total self-energy. 
To calculate this contribution, we ``approximate"  (at low $Q$)
the function $\xi_L(Q)$ 
with a Gaussian function of the form
\begin{equation}\label{xi_appr}
\xi_L^a(Q)= \alpha_L\exp\left(- {\frac {Q^2} {\tau_L^2}} \right) 
\end{equation}
 where the parameter $\tau_L$  is fixed in such way that 
the functions
$\xi_L^a(Q)$ and $\xi_L(Q)$
have the same Taylor expansion up to order $Q^2$.
The result for $\tau_L$ is the following:
\begin{equation}\label{tau_L}
\tau_L=\Lambda \left[ {\frac {1} {b ~\alpha_L} } \exp \left({\frac {1}  {b~ \alpha_L}} \right) 
\right]^{1/2}~.
\end{equation}
By using Eq. (\ref{xi_appr}) we calculate the corresponding self-energy in the standard way,
integrating up to infinity.
The total result for the self-energy calculated with the Technique B 
has the following analytic form:
\begin{equation}\label{vbar_2_b}
\bar V_V={\frac 4 3}  {\frac {1} {\sqrt{\pi}}} 
\left[ \left( \alpha_V- \alpha_L  \right)\tau +\alpha_L\tau_L  \right]~.
\end{equation}
The values of the parameters that reproduce $\alpha_S(Q)$ and fit the charmonium spectrum
are given in the columns (II B S) amd (II B M) of Tab. \ref{tabpar12},
for the scalar  and mass  interaction, respectively.
Note that the values of $\bar V_V$ obtained with Technique A and Technique B are very similar.
The behaviour of the obtained $\alpha_S(Q)$ is shown in Fig. \ref{fig4}.
We have checked that $\alpha_S(Q)$  of this Model II correctly gives the numerical values of
Eqs. (\ref{alphamz}) and (\ref{alphaq1}).

\vskip 0.5 truecm
\noindent
Summarizing the results of  this section, 
we have learned that a unique function, with the same definition
for all the values of $Q$, can reproduce the strong coupling constant of the 
quark-antiquark interaction.
This function is represented by the sum of a Gaussian plus a regularized
perturbative QCD function that gives a contribution also at low $Q$.
We shall take into accont this last point for the study of Model III.
The numerical results of Model II are similar to those of Model I
corroborating the physical validity of the two models.
The self-energy $\bar V_V$ for Model II can be calculated
in two different ways:
first, with the Technique A,
by introducing (analogously to what learned in  Model I), 
the quantity $\bar Q$ that, for this model, defines  the upper limit of the integration;
second, with the Technique B, we have also studied
an analytic procedure that ``extracts" the nonperturbative contribution
from the second term of the function that represents $\alpha_S(Q)$.
A procedure of the same kind well be used also in Model III.
As shown in Tab. \ref{tabpar12}, columns (II A S), (II A M),
(II B S) and (II B M), the numerical values of the parameters are similar,
demonstrating that the two Techniques A and B are almost equivalent.

\section{Model III. A differential equation for $\alpha_S(Q)$}\label{model3}
In this section  we obtain $\alpha_S(Q)$ as the solution of a (unique) differential
equation that represents the  low $Q$ and the high $Q$ physical behaviour
of the strong coupling costant.
This model is developed  taking  account of the results of the previous sections.

We start with some aspects related to notation and definitions.
\begin{itemize}
\item
For clarity in the calculation of the derivatives,
we prefer to use, in this section, the variable $t=Q^2$, in GeV$^2$.
\item
Only in the initial part of the model, we shall introduce the indices $l$ and $h$ 
for the low $t$
and high $t$ regions, respectively. For the strong coupling constant we have:
$\alpha_l(t)$ and $\alpha_h(t)$. 
The reader can assume  that, \textit{approximately},
the low $t$ region is defined by $t \leq \bar t$
and the high $t$ region by $t > \bar t$, being
$\bar t=\bar Q^2$,
where $\bar Q$ has been  introduced in the previous sections.
In the final form of the model, we shall consider simultaneously 
all the values of $t$ and drop those indices.
\item
To simplify the intermediate mathematical calculations, we also introduce:
$y_S(t)=1/\alpha_S(t)$, $y_l(t)=1/\alpha_l(t)$ and $y_h(t)=1/\alpha_h(t)$.
\item
Finally, we shall  consider, in some expressions, $y_S$, $y_l$ and $y_h$ as 
\textit{independent} variables and $t$ as a \textit{dependent} variable.  
\end{itemize}
We now study the differential  equations for the strong coupling constant
in the high $t$ and low $t$ regions, performing some transformations 
that allow to find a unique equation for the two cases.

We start from the low $t$ region, where,
as suggested by the results of the previous sections, 
  $\alpha_l(t)$  has essentially 
a Gaussian behaviour. 
It can be easily verified that  $\alpha_l(t)$ satisfies the following differential equation

\begin{equation}\label{diff_alpha_l}
{\frac {d \alpha_l(t)} {d t}}= -{ \frac {1} {\tau^2}} \alpha_l(t)
\end{equation}
The corresponding equation for $y_l(t)$ is:
\begin{equation}\label{diff_y_l}
{\frac {d y_l(t)} {d t}}= { \frac {1} {\tau^2}} y_l(t)
\end{equation}
Taking $y_l$ as an independent wariable, we can write equivalently
\begin{equation}\label{diff_t_l}
{\frac {d t} {d y_l}}= \tau^2  {\frac {1}  {y_l} }~.
\end{equation}
For completeness we write the solution of the previous equation in the form:
\begin{equation}\label{sol_diff_t_l}
t=\tau^2 \ln(y_l) +a_l
\end{equation}
being $a_l$ the integration constant.
With standard algebra one can verify that the Gaussian expression for $\alpha_l(t)$
is obtained:
\begin{equation}\label{sol_alpha_l}
\alpha_l(t)=\alpha_l(0) \exp \left( -{\frac {t} {\tau^2}}   \right)
\end{equation}
with
\begin{equation}\label{alpha_l_0}
\alpha_l(0)=\exp(a_l)~.
\end{equation}

We now consider the high $t$ region where the strong coupling constant
is given asymptotically by $\alpha_S^p(t)$ of Eq. (\ref{alpha_pert}).
For this region,
we have the following differential equation:
\begin{equation}\label{diff_alpha_h}
{\frac {d \alpha_h(t)} {d t}}= -{ \frac {b} { t}} \left[ \alpha_h(t) \right]^2
\end{equation} 
where $b$ is the usual constant introduced in Sect. \ref{running}~
with the numerical value given in Eq. (\ref{beta_lambda}) 
for our phenomenological calculation.
At fundamental level, the previous differential  equation represents 
the renormalization group equation
for $\alpha_S(Q)$ at the leading order \cite{pdg24,prs,chk}.

The corresponding equation for $y_h(t)$ is:
\begin{equation}\label{diff_y_h}
{\frac {d y_h(t)} {d t}}= { \frac {b} { t}}
\end{equation} 
Taking $y_h$ as an independent wariable, we can write equivalently
\begin{equation}\label{diff_t_h}
{\frac {d t} {d  y_h}}= {\frac  t b}
\end{equation} 
The solution has the form:
\begin{equation}\label{sol_diff_t_h}
t=a_h \exp({\frac {y_h} {b}})
\end{equation}
where $a_h$ is the integration constant.
To reproduce the experimental data in the perturbative regime, we set
\begin{equation}\label{a_h_def}
a_h=\Lambda^2
\end{equation}
where $\Lambda$ is the usual QCD scale introduced in Sect. \ref{running}~
with the numerical phenomenological value given in Eq. (\ref{beta_lambda}).
With standard passages, one can verify  that  $\alpha_h(t)$ takes  the form:
\begin{equation}\label{sol_alpha_h}
\alpha_h(t)=\alpha_S^p(t)
\end{equation}
where $\alpha_S^p(t)$ is 
the perturbative  expression  of Eq. (\ref{alpha_pert}).

Taking into account the
specific form of  Eqs. (\ref{diff_t_l}) and (\ref{diff_t_h}),
we can write a unique differential equation that holds
for all the values of $t$, that is:
\begin{equation}\label{diff_t_S}
{\frac {d t} {d  y_S}}= \tau^2  {\frac {1}  {y_S} } + {\frac t b}+ k~.
\end{equation} 
In the \textit{r.h.s.} of the previous equation,
the second term $ t/b$ is dominant at high $t$, 
while the first term contributes essentially at low $t$.
We have also introduced the additional constant term $k$, in GeV$^2$.
This term
will be related to the other parameters of Model III in  Eq. (\ref{k});
we shall also discuss in the following its physical relevance.
Here we note  that, in Eq. (\ref{diff_t_S}),
this constant term $k$ is negligible  at  high $t$; on the other hand, 
it does not alter too much the low $t$ behaviour that is given by the first term
of the previous equation.
We have used the symbol $\tau$ in the first term of Eq. (\ref{diff_t_S})
even though we shall not obtain a Gaussian term in the solution.
In this sense, $\tau$ has here a different meaning with respect to Model I and Model II.

In conclusion, Eq. (\ref{diff_t_S})  represents the ``unique" differential equation for
the strong coupling constant of our Model III.
The same equation can be rewritten as a differential equation
for $\alpha_S(t)$ in the following form:

\begin{equation}\label{diff_alpha_s}
{\frac {d \alpha_S(t)} {d  t}}= -\left [\alpha_S(t)\right ]^2 
{\frac {1} {\tau^2 \alpha_S(t)  + {\frac t b}+ k} }~.
\end{equation} 
We shall check that, with the solution for our model, 
the denominator 
in the \textit{r.h.s} of the previous equation
does not vanish and $\alpha_S(t)$  does not present any singularity.

The solution of Eq. (\ref{diff_alpha_s}) is easily found with the help of
of \textit{Maxima - A Computer Algebra System}, in the following implicit form:
\begin{equation}\label{sol_diff_t_s}
t+ k b=
\left[ 
a_S - \tau^2 \Gamma \left(0,{\frac {1} {b \alpha_S(t)} }\right) 
\right]
\exp \left( {\frac {1} {b \alpha_S(t)} } \right)~
\end{equation}
where it has been introduced the incomplete gamma function 
$\Gamma(0,x)$ whose properties can be found in Ref. \cite{grad}.
In particular, we recall
\begin{equation}\label{gaminc}
\Gamma(0,x)=\int_x^\infty {\frac {\exp(-v)} {v} }dv
\end{equation}
given by Eq. (8.350.1) of Ref.  \cite{grad}.
In consequence, one has for the derivative
\begin{equation}\label{dergaminc}
{\frac {d\Gamma(0,x)} {d x}} =- {\frac {\exp(-x)} {x} }~.
\end{equation}
By means of this last equation, one can easily check that 
the expression of Eq. (\ref{sol_diff_t_s}) really represents a solution of
Eq. (\ref{diff_t_S}).

The integration constant $a_S$ of Eq. (\ref{sol_diff_t_s})
 is determined by comparing that equation, at large $t$,
with Eq. (\ref{sol_diff_t_h}) and using the same definition given in Eq. (\ref{a_h_def}),
that is:
\begin{equation}\label{a_s_def}
a_S=\Lambda^2~.
\end{equation}
In this way, the correct high $t$ behaviour of $\alpha_S(t)$ is obtained.
In order to determine the constant parameter $k$,
we calculate Eq. (\ref{sol_diff_t_s}) at $t=0$.
Using Eq. (\ref{a_s_def}) and the definition  $\alpha_S(0)=\alpha_V$,
we obtain the following expression:
\begin{equation}\label{k}
k=\left[ \Lambda^2 - \tau^2 \Gamma \left(0,{\frac {1} {b \alpha_V} }\right) \right]
\exp \left( {\frac {1} {b\alpha_V} }\right) ~.
\end{equation}
We note that the parameter $k$ is related to $\alpha_V$. 
In this way, due to the presence of $k$, it is possible
to fix $\alpha_V$ at the  value $\alpha_V\simeq 2$ that is needed 
to reproduce the charmonium spectrum. 
In the following we shall see that $k$ takes a negative  value.
%
We observe that,  at low $Q$, $\alpha_S(Q)$ is not given exactly by a Gaussian
function of $Q$. In this respect, we recall that also
in Model I we had a Gaussian plus a constant term and in Model II we had a Gaussian plus
a regularized perturbative QCD function.

In summary,  Eq. (\ref{sol_diff_t_s}) represents the implicit solution of 
Eq. (\ref{diff_alpha_s});
the integration constant $a_S$ is given by Eq. (\ref{a_s_def}) and the parameter $k$
is given by Eq. (\ref{k}).
The only two free parameters of this model are $\alpha_V$ and $\tau$.

For the following calculations, we have to invert numerically Eq. (\ref{sol_diff_t_s})
to obtain $\alpha_S(t)$.

We now have to calculate the finite self-energy $\bar V_V$.
To this aim, we take advantage of the Model II, Technique B. 
We ``approximate", at low $t$, $\alpha_S(t)$, with the following function:
\begin{equation}\label{alpha_appr}
\alpha_S^a(t)=\alpha_V \exp \left ( -{\frac {t} {\tau_S^2}}  \right ).
\end{equation}
We also require  that $\alpha_S(t)$ (implicitly defined
in Eq. (\ref{sol_diff_t_s}) ) and $\alpha_S^a(t)$ 
have the same Taylor expansion up to orden $t=Q^2$.
With standard calculations, one finds that this condition is satisfied
when the parameter $\tau_S$ has the following form:
\begin{equation}
\tau_S= \left (  \tau^2 +  {\frac {k} {\alpha_V} } \right )^{\frac 1 2}~.
\end{equation}
Then, $\bar V_V$ is obtained, as in Eq. (\ref{vbardefexpl}), 
integrating $\alpha_S^a(t)$ up to infinity.
The result is:
 \begin{equation}\label{vbarvmod3}
\bar V_V= 
{\frac 4 3}  \alpha_V 
{\frac {\tau_S} {\sqrt{\pi}}}
 \end{equation}

\begin{figure}
{
  \includegraphics{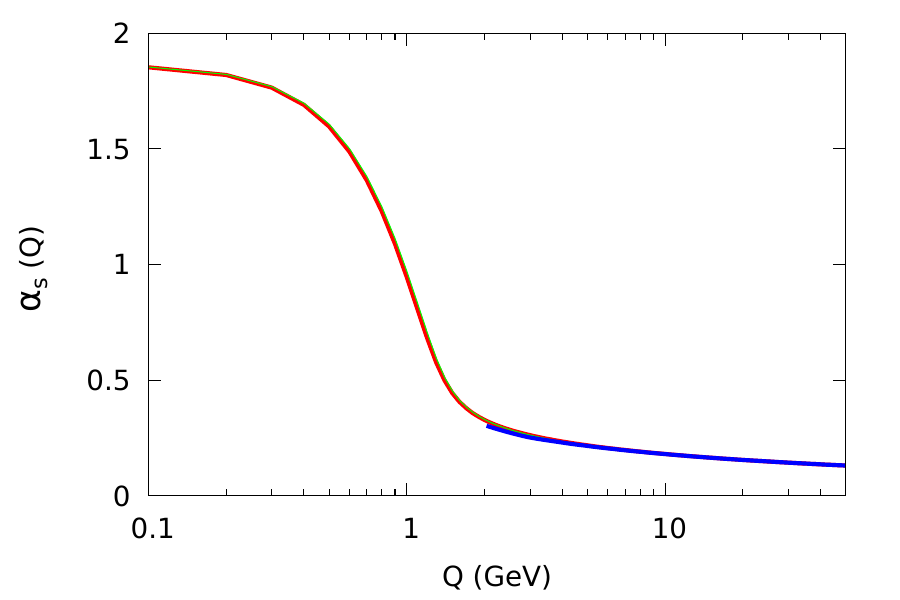}
}
\vspace{0.01cm}
\caption{ We plot $\alpha_S(Q)$ of
Model III for the scalar and mass interaction.
The definitions and  conventions are the same as in Fig. \ref{fig3}.
}
\label{fig5}       
\end{figure}

The values of the parameters that reproduce $\alpha_S(Q)$ and fit the charmonium spectrum
are given in the columns (III S) amd (III M) of Tab. \ref{tabpar3},
for the scalar  and mass  interaction, respectively.
The behaviour of the obtained $\alpha_S(Q)$ is shown in Fig. \ref{fig5}.
We have checked that $\alpha_S(Q)$ obtained by solving 
the differential equation of this model correctly gives the numerical values of
Eqs. (\ref{alphamz}) and (\ref{alphaq1}).

\section{The results of the charmonium spectrum}\label{charmspectr}
We give here some details about the reproduction of the charmonium spectrum
and comment the results of the calculations.
The technique for solving the RDLE of  Sect. \ref{dyn} 
and the procedure to fit the experimental data
\cite{pdg24}
are exactly the same  as in our previous works
\cite{localred,rednumb,scalint,effective}.
For the mass of the quark we have taken the  value $m_q=1.273$ GeV.
This value represents the  charm quark mass renormalized at the 
$\overline{MS}$ mass \cite{pdg24}.

\begin{figure}
{
  \includegraphics{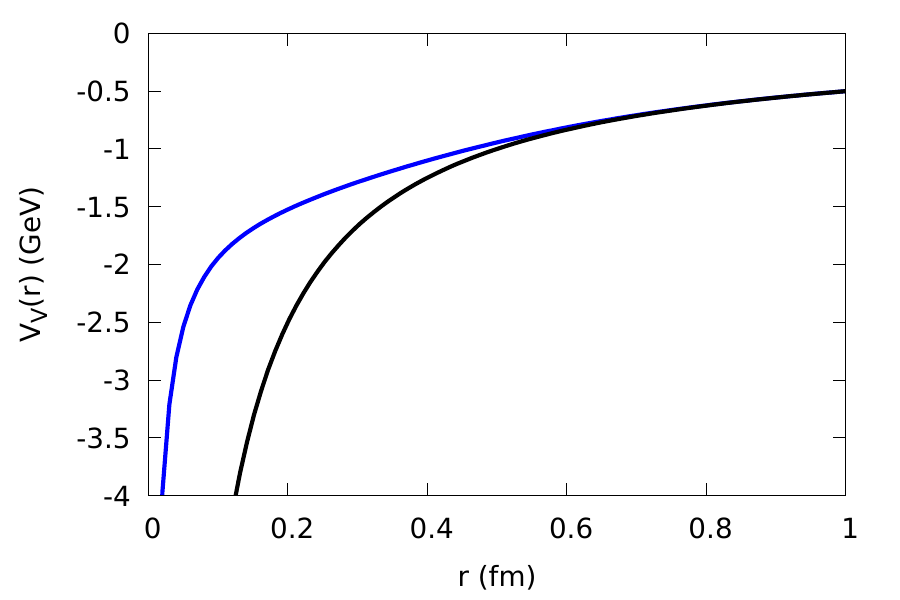}
}
\vspace{0.01cm}
\caption{
The blue line represents the vector potential function $V_V(r)$ for
the model II B M; the black line represents the Coulombic potential
with the same $\alpha_V$ of the Model II B M. 
The two functions coincide for large $r$. 
}
\label{fig_ii_b_m}       
\end{figure}

\begin{figure}
{
  \includegraphics{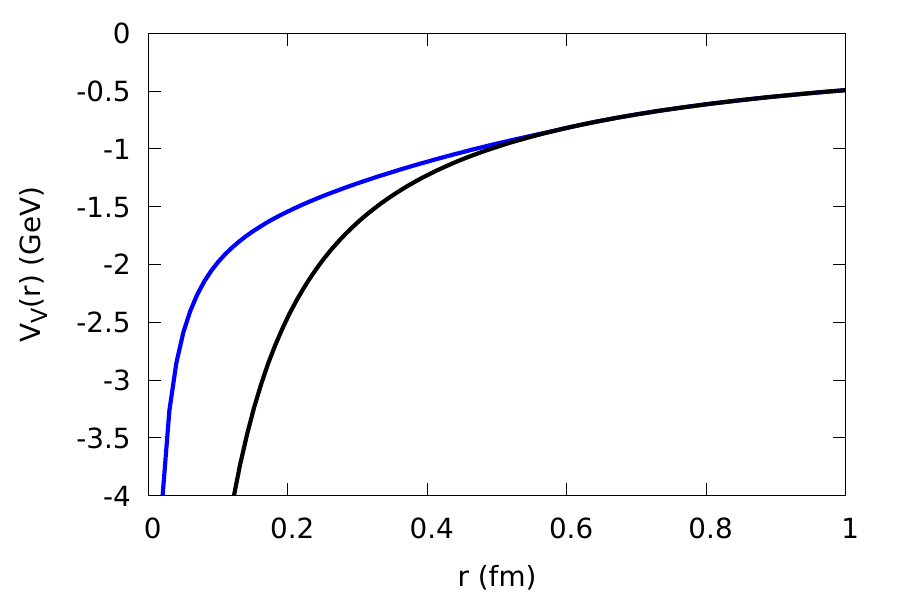}
}
\vspace{0.01cm}
\caption{
The vector function potential $V_V(r)$ for
the Model III S; the conventions are the same as in Fig.
\ref{fig_ii_b_m}.
}
\label{fig_iii_s}       
\end{figure}

The vector potential function  $V_V(r)$ is calculated performing numerically 
the Fourier transform of Eq. (\ref{potfourtrans}) for each model of $\alpha_S(Q)$.
The results obtained
for the different models 
 are very similar, so that we only show 
in Figs. \ref{fig_ii_b_m} and  \ref{fig_iii_s} the results 
for the Model II B M and III S, respectively. 
For comparison, we also plot, in the same figures, the pure Coulombic potential functions
observing that, at large $r$, our potential functions go to zero exactly as
the Coulombic ones.
Also the quark vector self-energy $\bar V_V$ is calculated according to the specifications
of each model.

The $X$-potential function 
$V_X(r)$ has the (negative) Gaussian form given in Eq. (\ref{vxgauss}).
The parameter $\bar V_X$ is determined, in this work, by means of Eq. (\ref{balance}).
On the other hand,
in Ref. \cite{scalint}, the  value of $\bar V_X$ was fixed
at $0.735$ GeV, phenomenologically related 
to the excitation of the first two scalar meson resonances that have 
the vacuum quantum numbers \cite{pdg24}: 
the $f_0(500)$ with the mass peak at (roughly) $M[ f_0(500)]=0.475$ GeV and
the $f_0(980)$ with the mass peak at (roughly) $M[ f_0(980)]=0.995$ GeV.
Being (indicatively) the mean value of these two mass peaks at $<M_0>=0.735$ GeV,
we took, in that work \cite{scalint}, $\bar V_X=<M_0> $.
In this work, $\bar V_X$ is  determined by Eq. (\ref{balance});
as shown in Tabs. \ref{tabpar12} and \ref{tabpar3}, 
it assumes values not far from $<M_0>$ and, in any case,
it belongs to the interval given by the first two scalar resonaces, that is:
$M[ f_0(500)] < \bar V_X < M[ f_0(980)] $.

\vskip 0.2 truecm
\noindent

For the quality of the fit, as in \cite{scalint,effective}, 
we have defined
\begin{equation}\label{qual}
\Theta=\sqrt{ {\frac {\sum_k(E_k^{th} -M_k^{exp} )^2} {N_d}} }~,
\end{equation}
where $E_k^{th}$ and $M_k^{exp}$ respectively represent
the result of the theoretical calculation and the experimental value 
of the mass,
for the $k$-th resonance and $N_d=16$ is the number of the fitted resonances.

The spectra obtained with the different models are very similar.
For this reason we only give in Tab. \ref{tabspectr} the results
of Model II B M and III S.

For completeness,
we also note that, as in \cite{scalint}, it is not possible to reproduce the resonance 
 $\chi_{c0}(3915)$. 
The new experimental data \cite{pdg24} give, for this resonance, a mass of
$3922.1 \pm 1.8 $ MeV.  
Our model, taking the quantum numbers $2^3P_0$,   
gives  the  mass  values of $3846$ MeV and $3857$ MeV, for the models II B M 
and  III S, respectively.  
Our model and other quark models give a wrong order for the masses of 
this resonance and its partner $\chi_{c1}(3872)$.
A possible solution of this problem has ben proposed by using three-body forces
in the framework of a phenomenological nonrelativistic model \cite{threebody}.

Concerning the general structure of our interaction, 
we note that  (neglecting the spin dependent terms
and recalling that $V_V(r)$ goes to zero as the Coulombic potential)
the maximum value of the \textit{total} vector  potential $\bar V_V +V_V(r)$ 
is given by $\bar V_V$.
In consequence, the maximum value for the mass of a bound state is, roughly,
$M_{max}=\bar V_V + 2 m_q $.
As shown in Tab. \ref{tabpar12}, $\bar V_V$ assumes slightly different values 
in the different models.
Taking, indicatively, $\bar V_V=1.83$ GeV,
we have the numerical value $M_{max}\approx 4.376 $ GeV.
Comparing this result with the analysis of the experimental data 
given in  Ref. \cite{pdg24}, we note that 
the only state with the properties of a conventional $c \bar c$ state 
and with a value of mass
\textit{greater than} $M_{max}$, would be the $\psi(4415)$.
This fact indicates that our value of $M_{max}$
(that  represents
the highest value of mass for which our  $c\bar c$ model can be safely applied) 
is approximately in accordance with the experimental findings.
At higher values of mass, new physical effects should be 
taken into account \cite{spectdec,santp,threebody,bramb24}
 and an explicit mechanism for confinement
should be introduced.

\vskip 0.2 truecm
\noindent
We conclude this paper with the following considerations.
The momentum dependence of $\alpha_S(Q)$ given by perturbative QCD, 
in accordance with high $Q$ experimental data, can be matched with the 
phenomenological behaviour of the same quantity at low $Q$.
By discarding an infinite contribution,
the vector quark \textit{finite} self-energy can be consistently calculated.
In this way  the charmonium spectrum is accurately reproduced.
Further investigation is necessary to establish a deeper connection between 
the effective bound state quark interaction and the phenomenology related 
to the QCD analysis.



\begin{table*} 
\caption{
Numerical values of the parameters of Model I and Model II; for more details
and for the definitions of the
parameters $\alpha_V, ~\tau, ~\alpha_G,~ \alpha_L,~ \bar Q$ and $\tau_L$
 see
Sect. \ref{model1}   and       Sect. \ref{model2}. The quark mass
$m_q $ is fixed at the value of Ref. \cite{pdg24}. 
$\bar V_V$ is the vector quark self-energy. 
For the scalar or mass interaction of Eq. (\ref{vxgauss}), the parameter $\bar V_X$ is determined by
Eq. (\ref{balance}), 
$r_X$ is a fit parameter.
The parameter
$\Theta$, defined in Eq. (\ref{qual}), gives the quality of the fit.
}
\begin{center}
\begin{tabular}{lllllllll}
\hline 
\hline \\   
            &           &          &           &         &           &           & Units \\ 
\hline \\
 $m_q $      &  $1.273$   &         &           &         &           &           & GeV   \\
$\Lambda$    &  $0.1462$ &         &           &         &           &           & GeV   \\
$b$         & $0.65842$   &         &           &         &           &           &      \\

\hline \\
Model       &   I S    & I M      & II A S      & II A M     & II B S     & II B M   & \\     
\hline \\   
$\alpha_V $ &  $1.9036 $ & $1.9740$  & $1.9010  $ & $1.9042 $  &  $1.8997  $ & $1.8994 $ &  \\
$\tau     $ &  $1.0405 $ & $1.0361$  & $1.0199  $ & $1.0238 $  &  $1.0196  $ & $1.0204 $ & GeV\\
$\alpha_G $ &  $1.6401 $ & $1.7105$  &           &           &            &          &   \\
$\alpha_L $ &           &           & $0.39802 $ & $0.44694$ &  $0.37300 $ & $0.36462 $&  \\  
$\bar Q   $ &  $2.3608$ & $2.3629$   & $2.3379  $ & $2.4496 $  &            &          & GeV \\
$\tau_L   $ &           &           &           &           &  $2.2599  $ & $2.3951  $& GeV \\
$\bar V_V $ &  $1.8100 $ & $1.8600$   & $1.8045  $ & $1.8340$  &  $1.80505 $ & $1.8350 $& GeV \\
$\bar V_X $ &  $0.73000$ & $0.68000$  & $0.73546 $ & $0.70596$ &  $0.73495 $ & $0.70496 $& GeV\\
$r_X      $ &  $1.8437 $ & $2.0041$   & $1.8022  $ & $1.8718 $  &  $1.8097  $ & $1.8894  $& fm\\
\hline \\
$\Theta  $  &  $12.5  $ & $14.0 $   & $12.9   $ & $12.2  $  &  $12.9   $ & $12.0   $& MeV \\
\hline\\
\end{tabular}
\end{center}

\label{tabpar12}
\end{table*}

\begin{table*} 
\caption{
Numerical values of the parameters of Model III.
The values of the quark mass
$m_q $, $\Lambda$ and $b$, not shown here,
 are \textit{exactly} the same as in Tab. \ref{tabpar12}.
 For more details,
in particular for the definitions of $\tau,~k$ and $\tau_S$,
 see Sect. \ref{model3}.
For the meaning of the other parameters also see Tab. \ref{tabpar12}.
}
\begin{center}
\begin{tabular}{lllll}
\hline 
\hline \\   
            &           &             & Units \\ 
\hline \\
Model        &   III S   & III M       &  \\      
\hline \\   
$\alpha_V $ &  $1.8647 $ & $1.8659 $    &     \\
$\tau     $ &  $1.9146 $ & $1.9501 $   & GeV \\
$ k       $ &  $-3.7306$ & $-3.8781 $   & GeV$^2$    \\
$\tau_S   $ &  $1.2904 $ & $1.3145  $  
 &     GeV \\
$\bar V_V $ &  $1.8100 $ & $1.8450$   &     GeV \\
$\bar V_X $ &  $0.729955 $ & $0.69503$   &     GeV \\
$r_X      $ &  $1.8732 $ & $1.9852$   &      fm \\
\hline \\
$\Theta  $  &  $12.55  $ & $14.01 $   &    MeV \\
\hline\\
\end{tabular}
\end{center}

\label{tabpar3}
\end{table*}

\begin{table*}

\caption{
Comparison between the  theoretical results of the charmonium spectrum
and the experimental average values \cite{pdg24}, listed in the last column.
As examples,
we only show (in the corresponding columns) the results of the models
 II M B and III S, respectively studied in Sects. \ref{model2} and   \ref{model3}.
The  values of the parameters used for these models can be found
 in Tabs. \ref{tabpar12} and \ref{tabpar3}.
The quantum numbers  $n$, $L$, $S$ and $J$, introduced in Ref. \cite{rednumb},
respectively
represent the principal quantum number, the orbital angular momentum, the spin 
and the total  angular momentum.
A line divides the resonances below and above the open Charm threshold.
  }

\begin{center}
\begin{tabular}{ccccc}
\hline
\hline \\
Name & $n^{2S+1}L_J$  & II M B &  III S & Experiment          \\
\hline \\
$\eta_c(1S)$    &  $1^1 S_0 $     & 2983   & 2984   & 2984.1   $\pm$  0.4   \\
$J/\psi(1S)$    &  $1^3 S_1 $     & 3113   & 3099   & 3096.9   $\pm$  0.006 \\
$\chi_{c0}(1P)$ &  $1^3 P_0 $     & 3403  & 3420   & 3414.71  $\pm$ 0.30   \\
$\chi_{c1}(1P)$ &  $1^3 P_1 $     & 3497  & 3503   & 3510.67  $\pm$ 0.05   \\
$ h_c(1P)$      &  $1^1 P_1 $     & 3514  & 3516    & 3525.37  $\pm$ 0.14   \\ 
$\chi_{c2}(1P)$ &  $1^3 P_2 $     & 3572   & 3565    & 3556.17  $\pm$ 0.07   \\
$\eta_c(2S)$    &  $2^1 S_0 $     & 3630   & 3638    & 3637.7   $\pm$ 0.9    \\
$\psi(2S)$      &  $2^3 S_1 $     & 3681  & 3679    & 3686.097 $\pm$0.011\\
\\
\hline \\
$\psi(3770)$&     $1^3 D_1 $     &  3790 & 3797  & 3773.7   $\pm$ 0.7 \\  
$\psi_2(3823)$&   $1^3 D_2 $     &  3827 & 3829  & 3823.51   $\pm$ 0.34  \\
$\chi_{c1}(3872)$&$2^3 P_1 $     &  3890 & 3895  & 3871.64  $\pm$ 0.06 \\
$\chi_{c2}(3930)$&$2^3 P_2 $     &  3933 & 3929  & 3922.5   $\pm$ 1.0  \\
$\psi(4040)$&     $3^3 S_1 $     &  4015 & 4013  & 4040    $\pm$ 4     \\
%
$\chi{c1}(4140)$& $3^3 P_1 $     &  4147 & 4145 & 4146.5  $\pm$ 3.0    \\
$\psi(4230)    $& $4^3 S_1 $     &  4221 & 4211  & 4222.1   $\pm$ 2.3     \\
$\chi{c1}(4274)$& $4^3 P_1 $     &  4287 & 4269  & 4286    $\pm$ 9      \\
\\
\hline
\hline \\
~\\
~\\
~\\
~\\

\end{tabular}
\end{center}
\label{tabspectr}
\end{table*}



\vskip 10.0 truecm
\newpage

\begin{thebibliography}{50}  
%
\bibitem{chromomds}
M. De Sanctis, Front. Phys. {\bf 7}, 25 (2019).

\bibitem{localred}
M. De Sanctis, Acta Phys. Pol. B {\bf 52}, 125 (2021).
\bibitem{rednumb}
M. De Sanctis, Acta Phys. Pol. B {\bf 52}, 1289 (2021).
\bibitem{relvar}
M. De Sanctis, Acta Phys. Pol. B {\bf 53}, 7A-2 (2022).
\bibitem{scalint}
M. De Sanctis, Acta Phys. Pol. B {\bf 54}, 1A-2 (2023).
\bibitem{effective}
M. De Sanctis, Acta Phys. Pol. B {\bf 54}, 10A-3 (2023).

\bibitem{Deur05}
A. Deur, V. Burkert, J.P. Chen and W. Korsch,
``Experimental determination of the effective strong coupling constant'', 
Phys.Lett. B {\bf 650}, 244 (2007);
arXiv:hep-ph/0509113v3, (2007).
\bibitem{Deur08}
A. Deur, V. Burkert, J.P. Chen and W. Korsch,
``Determination of the effective strong coupling constant 
$\alpha_{s,g1}(Q^2)$
from CLAS spin structure function data'',
Phys.Lett.B {\bf 665}, 349 (2008);
arXiv:hep-ph/0803.4119v2, (2008).
\bibitem{Deur22}
A. Deur, V. Burkert, J.P. Chen and W. Korsch,
``Experimental determination of the QCD effective charge
$\alpha_{g1}(Q)$'', Particles {\bf 5(2)}, 171 (2022);
arXiv:2205.01169v2 [hep-ph] (2022).
\bibitem{Deurrev16}
A. Deur, S. J. Brodsky and G. F. de T\'eramond,
`` The QCD running coupling'',
Progress in Particle and Nuclear Physics, {\bf 90}, 1 (2016).
\bibitem{Deurrev23}
A. Deur, S. J. Brodsky and C. D. Roberts,
``QCD Running Couplings and Effective Charges''
arXiv:2303.00723v1 [hep-ph] (2023), 
Review commissioned by 
Progress in Particle and Nuclear Physics.
\bibitem{coulreg}
 Shwe Sin Oo, Sungwook Lee and Khin Maung Maung,
``Modified Coulomb potential and the S-state wavefunction of heavy quarkonia''
arXiv:2406.06790v1 [hep-ph] (2024).
\bibitem{hlfmatch}
H. Cancio and P. Masjuan,
``The Holographic QCD Running Coupling Constant from the Ricci Flow''
 	arXiv:2408.00455 [hep-ph] (2024).


\bibitem{richar79}
J. L. Richardson, Phys. Lett. B {\bf 82}, 272 (1979).

\bibitem{spectdec}
Chaitanya Anil Bokade and Bhaghyesh,
``Charmonium: Conventional and XYZ States in a Relativistic Screened Potential Model'',
 arXiv:2408.06759v1 [hep-ph] (2024).

\bibitem{santp}
J. Ferretti and E. Santopinto.
``Quark structure of the X(4500), X(4700) and $\chi_c(4P,5P)$ states'',
 arXiv:2104.00918v1 [hep-ph] (2021).

\bibitem{threebody}
Sungsik Noh, Aaron Park, Hyeongock Yun, Sungtae Cho and Su Houng Lee,
``The Inevitable Quark Three-Body Force and its Implications for Exotic States'',
 arXiv:2408.00715v1 [hep-ph] (2024).

\bibitem{bramb24}
M. Berwein, N. Brambilla, A. Mohapatra and A. Vairo
``One Born-Oppenheimer Effective Theory to rule them all: hybrids, tetraquarks, pentaquarks, doubly heavy baryons and quarkonium'',
 arXiv:2408.04719v1 [hep-ph] (2024).






\bibitem{pdg24}
S. Navas et al. (Particle Data Group), Phys. Rev. D 110, 030001 (2024)


\bibitem{prs}
 G. M. Prosperi, M. Raciti and C. Simolo,
Prog. Part. Nucl. Phys.  {\bf 58} 387 (2007), arXiv:hep-ph/0607209. 

\bibitem{chk}
J. Campbell, J. Huston, F. Krauss, 
The Black Book of Quantum Chromodynamics,
A Primer for the LHC Era,
Oxford University Press, Oxford U.K.(2018). 


\bibitem{grad}
I.S. Gradshteyn and I.M. Ryzhik, ``Table of integrals, series and products'', Academic
Press, New York, (1980).



%

%

%

\end{thebibliography}
\end{document}